\documentclass{emulateapj}
\usepackage{apjfonts}
\usepackage{rotating}

\newcommand{\pivec}{\mbox{\boldmath $\pi$}}
\newcommand{\xivec}{\mbox{\boldmath $\xi$}}

\lefthead{HAN ET AL.} 
\righthead{MICROLENSING PLANET AROUND BROWN-DWARF}

\begin{document}
\title{
Microlensing Discovery of a Tight, Low Mass-ratio Planetary-mass Object around an Old, Field Brown Dwarf}

\author{
C. Han$^{1\ast}$, 
Y. K. Jung$^{1}$, 
A. Udalski$^{O1\dag}$, 
T. Sumi$^{M1\ddag}$, 
B. S. Gaudi$^{U1\ast}$, 
A. Gould$^{U1\ast}$, 
D. P. Bennett$^{M10\ddag}$, 
Y. Tsapras$^{R1,R2\clubsuit}$,\\
and\\
M. K. Szyma\'nski$^{O1}$, 
M. Kubiak$^{O1}$, 
G. Pietrzy\'nski$^{O1,O2}$, 
I. Soszy\'nski$^{O1}$, 
J. Skowron$^{O1}$, 
S. Koz{\l}owski$^{O1}$, 
R. Poleski$^{O1,U1}$, 
K. Ulaczyk$^{O1}$, 
\L. Wyrzykowski$^{O1,O3}$, 
P. Pietrukowicz$^{O1}$, \\
(The OGLE Collaboration),\\
F. Abe$^{M2}$, 
I. A. Bond$^{M3}$, 
C. S. Botzler$^{M4}$, 
P. Chote$^{M5}$, 
M. Freeman$^{M4}$, 
A. Fukui$^{M6}$, 
K. Furusawa$^{M2}$,
P. Harris$^{M5}$, 
Y. Itow$^{M2}$, 
C. H. Ling$^{M3}$, 
K. Masuda$^{M2}$, 
Y. Matsubara$^{M2}$, 
Y. Muraki$^{M2}$, 
K. Ohnishi$^{M7}$, 
N. J. Rattenbury$^{M4}$, 
To. Saito$^{M8}$, 
D. J. Sullivan$^{M5}$, 
W. L. Sweatman$^{M3}$, 
D. Suzuki$^{M1}$, 
P. J. Tristram$^{M9}$, 
K. Wada$^{M1}$, 
P. C. M. Yock$^{M4}$,\\
(The MOA Collaboration),\\
V. Batista$^{U1}$, 
G. Christie$^{U6}$, 
J.-Y. Choi$^{1}$, 
D. L. DePoy$^{U2}$, 
Subo Dong$^{U3}$, 
K.-H. Hwang$^{1}$, 
A. Kavka$^{U1}$, 
C.-U. Lee$^{U4}$, 
L. A. G. Monard$^{U5}$, 
T. Natusch$^{U6}$, 
H. Ngan$^{U6}$, 
H. Park$^{1}$, 
R. W. Pogge$^{U1}$, 
I. Porritt$^{U7}$, 
I.-G. Shin$^{1}$, 
T.G. Tan$^{U8}$, 
J. C. Yee$^{U1}$, \\
(The $\mu$FUN Collaboration), \\
K. A. Alsubai$^{R7}$, 
V. Bozza$^{R8}$,
D. M. Bramich$^{R3}$, 
P. Browne$^{R4}$, 
M. Dominik$^{R4}$, 
K. Horne$^{R4}$, 
M. Hundertmark$^{R4}$, 
S. Ipatov$^{R7}$, 
N. Kains$^{R3}$, 
C. Liebig$^{R4}$, 
C. Snodgrass$^{R6}$, 
I. A. Steele$^{R5}$, 
R. A. Street$^{R1}$\\
(The RoboNet Collaboration)\\
}

\bigskip\bigskip

\affil{$^{1\ast}$Department of Physics, Chungbuk National University, Cheongju 371-763, Republic of Korea}
\affil{$^{O1}$Warsaw University Observatory, Al. Ujazdowskie 4, 00-478 Warszawa, Poland} 
\affil{$^{O2}$Universidad de Concepci\'{o}n, Departamento de Astronomia, Casilla 160-C, Concepci\'{o}n, Chile} 
\affil{$^{O3}$Institute of Astronomy, University of Cambridge, Madingley Road, Cambridge CB3 0HA, UK} 
\affil{$^{M1}$Department of Earth and Space Science, Osaka University, Osaka 560-0043, Japan}
\affil{$^{M2}$Solar-Terrestrial Environment Laboratory, Nagoya University, Nagoya, 464-8601, Japan} 
\affil{$^{M3}$Institute of Information and Mathematical Sciences, Massey University, Private Bag 102-904, North Shore Mail Centre, Auckland, New Zealand} 
\affil{$^{M4}$Department of Physics, University of Auckland, Private Bag 92-019, Auckland 1001, New Zealand}
\affil{$^{M5}$School of Chemical and Physical Sciences, Victoria University, Wellington, New Zealand} 
\affil{$^{M6}$Okayama Astrophysical Observatory, National Astronomical Observatory of Japan, Asakuchi, Okayama 719-0232, Japan} 
\affil{$^{M7}$Nagano National College of Technology, Nagano 381-8550, Japan} 
\affil{$^{M8}$Tokyo Metropolitan College of Aeronautics, Tokyo 116-8523, Japan}
\affil{$^{M9}$Mt. John University Observatory, P.O. Box 56, Lake Tekapo 8770, New Zealand} 
\affil{$^{M1}$University of Notre Dame, Department of Physics, 225 Nieuwland Science Hall, Notre Dame, IN 46556-5670, USA} 
\affil{$^{U1}$Department of Astronomy, Ohio State University, 140 West 18th Avenue, Columbus, OH 43210, USA}
\affil{$^{U2}$Department of Physics and Astronomy, Texas A\&M University, College Station, TX 77843, USA} 
\affil{$^{U3}$Institute for Advanced Study, Einstein Drive, Princeton, NJ 08540, USA}
\affil{$^{U4}$Korea Astronomy and Space Science Institute, Daejeon 305-348, Republic of Korea} 
\affil{$^{U5}$Klein Karoo Observatory, Calitzdorp, and Bronberg Observatory, Pretoria, South Africa}
\affil{$^{U6}$Auckland Observatory, Auckland, New Zealand}
\affil{$^{U7}$Turitea Observatory, Palmerston North, New Zealand}
\affil{$^{U8}$Perth Exoplanet Survey Telescope, Perth, Australia}
\affil{$^{R1}$Las Cumbres Observatory Global Telescope Network, 6740B Cortona Dr, Goleta, CA 93117, USA} 
\affil{$^{R2}$School of Physics and Astronomy, Queen Mary University of London, Mile End Road, London E1 4NS, UK} 
\affil{$^{R3}$European Southern Observatory, Karl-Schwarzschild-Str. 2, 85748 Garching bei M??nchen, Germany} 
\affil{$^{R4}$SUPA, School of Physics \& Astronomy, University of St Andrews, North Haugh, St Andrews KY16 9SS, UK}
\affil{$^{R5}$Astrophysics Research Institute, Liverpool John Moores University, Liverpool CH41 1LD, UK}
\affil{$^{R6}$Max Planck Institute for Solar System Research, Max-Planck-Str. 2, 37191 Katlenburg-Lindau, Germany}
\affil{$^{R7}$Alsubai Establishment for Scientific Studies, Doha, Qatar}
\affil{$^{R8}$Universit\`{a} degli Studi di Salerno, Dipartimento di Fisica ``E.R. Caianiello'', Via S. Allende, 84081 Baronissi (SA), Italy}
\affil{$^{\ast}$The $\mu$FUN Collaboration} 
\affil{$^{\dag}$The OGLE Collaboration}
\affil{$^{\ddag}$The MOA Collaboration} 
\affil{$^{\clubsuit}$The RoboNet Collaboration}

\begin{abstract}
Observations of accretion disks around young brown dwarfs have led to the 
speculation that they may form planetary systems similar to normal stars. 
While there have been several detections of planetary-mass objects around 
brown dwarfs (2MASS 1207-3932 and 2MASS 0441-2301), these companions have 
relatively large mass ratios and projected separations, suggesting that 
they formed in a manner analogous to stellar binaries. We present the 
discovery of a planetary-mass object orbiting a field brown dwarf via g
ravitational microlensing, OGLE-2012-BLG-0358Lb. The system is a low 
secondary/primary mass ratio (0.080 $\pm$ 0.001), relatively tightly-separated 
($\sim 0.87$ AU) binary composed of a planetary-mass object with 1.9 $\pm$ 
0.2 Jupiter masses orbiting a brown dwarf with a mass 0.022 $M_\odot$. 
The relatively small mass ratio and separation suggest that the companion 
may have formed in a protoplanetary disk around the brown dwarf host, in 
a manner analogous to planets.
\end{abstract}


\section{Introduction}

Brown dwarfs (BDs) are sub-stellar objects that are too low in mass to 
sustain hydrogen fusion reactions in their cores. Although still a matter 
of debate, the most popular theory about the origin of BDs is that they 
form via direct collapse similar to stars, perhaps aided by turbulent 
fragmentation (see \citet{luhman12} for a review). This theory is supported 
by observational evidence showing that several medium-sized BDs 
are girdled by disks of material \citep{luhman05, apai05, ricci12}. 
The existence of accretion disks around these failed stars naturally 
leads to the speculation that BDs may also harbor planetary 
systems analogous to those found in abundance around stars.

There have been several detections of planetary-mass objects around brown 
dwarfs: 2MASS 1207-3932B with $M_{\rm B}\sim 4 M_{\rm J}$ \citep{chauvin04} and 2MASS 
0441-2301B with $M_{\rm B}\sim 7.5 M_{\rm J}$ \citep{todorov10}.  However, these systems 
have relatively large mass ratios of $q\sim 0.16$ for 2MASS 1207-3932 and 
$q\sim 0.25$ -- 0.5 for 2MASS 0441-2301, more akin to binary stellar systems. 
Furthermore, they have relatively large separations ($\sim 15$ AU for 2MASS 
0441-2301 and $\sim 45$ AU for 2MASS 1207-3932), likely near or beyond the 
outer edges of the accretion disks observed around BDs 
\citep{luhman07, ricci12, ricci13}.  Therefore, it seems unlikely that these 
companions formed from the protoplanetary disk material via either of the 
popular giant planet formation mechanisms of core accretion \citep{pollack96} 
or disk fragmentation \citep{kuiper51, cameron78, boss97, durisen07}.
 Rather, these are more likely to have formed like stellar binaries, through 
the process of gravitational fragmentation of massive primordial disk 
\citep{lodato05}. Thus, according to a classification system based on their 
formation, they are not bona fide planets.

\begin{figure*}[ht]
\epsscale{0.8}
\plotone{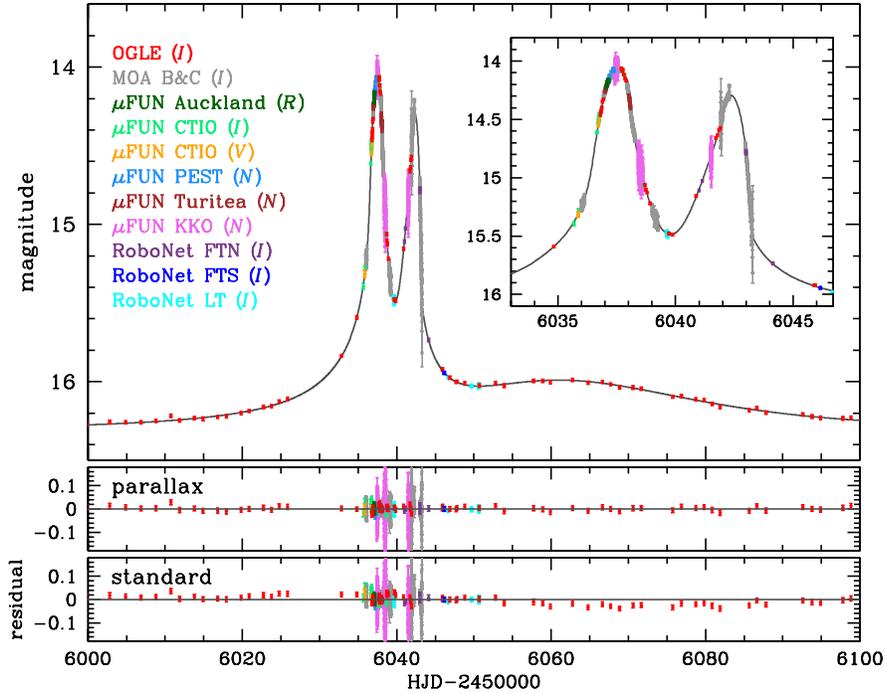}
\caption{\label{fig:one}
Light curve the microlensing event OGLE-2012-BLG-0358. The lower 2 panels 
show the residuals from the best-fit standard binary-lens model and from 
the model considering the parallax effect. The letters after the individual 
telescopes represent the pass bands of observation.
}\end{figure*}

In this paper, we report a microlensing discovery of a tight, low mass-ratio 
planetary-mass object orbiting an old, field BD that we suggest may have 
formed in a protoplanetary disk. Microlensing is the astronomical phenomenon 
wherein the brightness of a star is magnified by the bending of light due to 
the gravity of an intervening object (lens) positioned between the background 
star (source) and an observer. This discovery was possible, in spite of the 
extremely low luminosity of the BD, because the lensing phenomenon occurs 
regardless of the lens brightness.

\section{Observation}

The BD planetary system was discovered in the microlensing event OGLE-2012-BLG-0358. 
The event occurred in 2012 observing season on a star located in the Galactic Bulge 
field with equatorial coordinates $({\rm RA}, {\rm DEC})_{2000} = (17^{\rm h}
42^{\rm m}46.77^{\rm s}, -24^\circ 15'39.6'')$, which corresponds to the Galactic 
coordinates $(l, b)_{2000} = (3.65^\circ, 2.99^\circ)$. It was first discovered 
by the Optical Gravitational Lensing Experiment (OGLE: \citet{udalski03}) group 
in April 2012. During its early phase, the light curve of the event appeared to 
be a high-magnification event produced by a single mass. Since high-magnification 
events are prime targets for planet detections, the event was additionally observed by 
other groups including the Microlensing Follow-Up Network ($\mu$FUN: \citet{gould06}), 
Microlensing Observations in Astrophysics (MOA: \citet{bond01, sumi03}), and RoboNet 
\citep{tsapras09}. As the event approached its peak, it was noticed that the light 
curve deviated from a standard single-lens light curve and the anomaly became obvious 
as the light curve peaked again ~5 days after the first peak. Continued observations 
by the OGLE group revealed that the event produced another extended weak bump. In 
Table~\ref{table:one}, we list the telescopes used for observation.

In Figure~\ref{fig:one}, we present the light curve of the event. It is characterized 
by two strong peaks centered at Heliocentric Julian Date (HJD)$\sim$2456537.5 and 
2456542.5 and an extended weak bump centered at HJD~2456065. A strong peak in a 
lensing light curve occurs when a source star approaches close to or crosses the 
tip of a caustic produced by a lens composed of multiple objects. The caustic 
represents the envelope of light rays refracted by a curved surface and it is 
commonly visible as a curved region of bright light appearing when light shines 
on a drinking glass. For a gravitational lens composed of two masses, caustics 
form a single or multiple sets of closed curves each of which is composed of 
concave curves that meet at cusps.

\begin{deluxetable}{ll}
\tablecaption{Telescopes\label{table:one}}
\tablewidth{0pt}
\tablehead{
\multicolumn{1}{c}{group} &
\multicolumn{1}{c}{telescope}
}
\startdata
OGLE       & 1.3 m Warsaw, Las Campanas, Chile                         \\
MOA        & 0.6 m Boller \& Chivens, Mt. John, New Zealand            \\
$\mu$FUN   & 1.3 m SMARTS, Cerro Tololo Inter-American (CTIO), Chile   \\
$\mu$FUN   & 0.4 m Auckland, New Zealand                               \\
$\mu$FUN   & 0.36 m Klein Karoo Observatory (KKO), South Africa        \\ 
$\mu$FUN   & 0.3 m Perth Extrasolar Survey Telescope (PEST), Australia \\
$\mu$FUN   & 0.4 m Turitea, New Zealand                                \\
RoboNet    & 2.0 m Faulkes North Telescope (FTN), Hawaii, USA          \\
RoboNet    & 2.0 m Faulkes South Telescope (FTS), Australia            \\
RoboNet    & 2.0 m Liverpool Telescope (LT), Canary Islands, Spain     
\enddata  
\end{deluxetable}

\begin{figure}[ht]
\epsscale{1.2}
\plotone{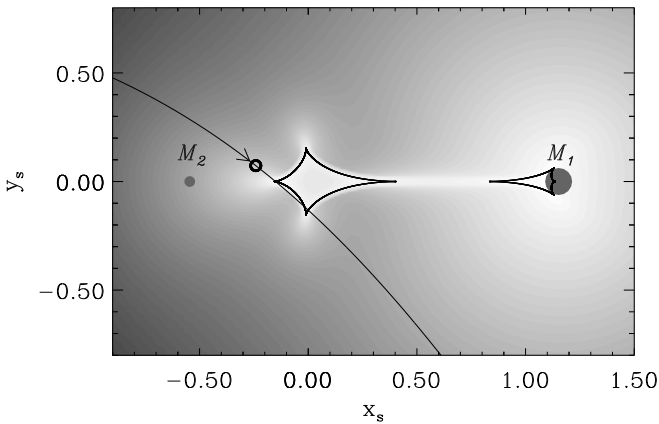}
\caption{\label{fig:two}
Geometry of the lens system. The closed figures composed concave curves 
represent the caustic and the line with an arrow is the source trajectory. 
$M_1$ and $M_2$ represent the binary lens components, where M1 is the heavier 
one. Greyscale represents the lensing magnification where brighter tone 
denotes higher magnifications. All lengths are scaled by the Einstein radius 
corresponding to the total mass of the binary lens.
}\end{figure}

\section{Modeling}

With the signature of lens multiplicity, we conduct binary-lens modeling of the 
observed light curve. Basic description of a binary-lens light curve requires 7 
lensing parameters. Three of these parameters describe the lens-source approach, 
including the time of the closest source approach to a reference position of the 
binary lens, $t_0$, the separation between the source and the reference position, 
$u_0$ (normalized by the angular Einstein radius $\theta_{\rm E}$), and the time 
scale for the source to cross $\theta_{\rm E}$ (Einstein time scale $t_{\rm E}$). The Einstein 
ring denotes the image of a source in the event of perfect lens-source alignment, 
and its radius is commonly used as the length scale of lensing phenomena. Another 
three lensing parameters describe the binary nature of the lens, including the 
projected separation, $s$ (normalized by $\theta_{\rm E}$), and the mass ratio, $q$, 
between the binary components, and the angle between the source trajectory and the 
binary axis, $\alpha$ (source-trajectory angle). The last parameter is the normalized 
source radius $\rho_*=\theta_*/\theta_{\rm E}$, where $\theta_*$ is the angular source 
radius. This parameter is needed to precisely describe the parts of a lensing light 
curve involved with caustic crossings or approaches of the source during which the 
lensing light curve is affected by the finite size of the source star. In our modeling 
of finite-source effects, we additionally consider the limb-darkening variation of 
the source star surface by modeling the surface brightness profile as a standard 
linear law.

We search for a solution of lensing parameters that best describes the observed 
light curve by minimizing $\chi^2$ in the parameter space encompassing wide ranges 
of binary separations and mass ratios. For $\chi^2$ minimization, we use the Markov 
Chain Monte Carlo method. In order to properly combine data sets obtained from 
different observatories, we readjust photometric errors of the individual data 
sets first by adding a quadratic error term so that the cumulative distribution 
of $\chi^2$ ordered by magnifications matches to a standard cumulative distribution 
of Gaussian errors and then by rescaling errors so that $\chi^2$ per degree of freedom 
becomes unity for each data set. We eliminate data points with large errors and 
obvious outlyers to minimize their effect on modeling.

From the initial search for solutions obtained from modeling based on the standard 
binary-lensing parameters (standard model), we find a solution of a binary lens with 
a projected separation $s\sim 1.7$ and a mass ratio $q\sim 9.8$. See 
Table~\ref{table:two} for the complete solution. Although the model describes the 
main feature of the two strong peaks, it is found that there exist long-term residuals 
in the wings of the light curve including the extended weak bump as shown in the bottom 
panel of Figure~\ref{fig:one}. This suggests the need to consider higher-order effects.

\begin{deluxetable*}{lrrrrr}
\tablecaption{Lensing Parameters\label{table:two}}
\tablewidth{0pt}
\tablehead{
\multicolumn{1}{l}{parameters} &
\multicolumn{5}{c}{model} \\
\multicolumn{1}{c}{}                       &
\multicolumn{1}{c}{standard           }    &
\multicolumn{1}{c}{parallax ($u_0>0$) }    &
\multicolumn{1}{c}{parallax ($u_0<0$) }    &
\multicolumn{1}{c}{orbit + parallax   }    &
\multicolumn{1}{c}{xallarap ($P=1$ yr)}    
}
\startdata
$\chi^2/{\rm dof}$        & 2347.81/1592      & 1598.22/1590      & 1596.24/1590      &  1595.17/1588      & 1601.53/1588      \\
$t_0$ (HJD')              & 6040.24$\pm$0.01  & 6040.33$\pm$0.01  & 6040.33$\pm$0.01  &  6057.49$\pm$0.10  & 6040.33$\pm$0.01  \\
$u_0$                     & 0.108$\pm$0.001   & 0.098$\pm$0.001   & -0.098$\pm$0.001  &  -0.832$\pm$0.002  & 0.098$\pm$0.001   \\
$t_{\rm E}$ (days)        & 24.38$\pm$0.07    & 26.47$\pm$0.11    & 26.46$\pm$0.11    &  25.64$\pm$0.08    & 26.46$\pm$0.12    \\
$s_1$                     & 1.687$\pm$0.002   & 1.696$\pm$0.003   & 1.696$\pm$0.003   &  1.700$\pm$0.002   & 1.696$\pm$0.003   \\
$q_1$ ($10^{-2}$)         & 9.810$\pm$0.071   & 12.531$\pm$0.154  & 12.486$\pm$0.159  &  12.281$\pm$0.118  & 12.487$\pm$0.179  \\
$\alpha$                  & 5.544$\pm$0.001   & -0.721$\pm$0.001  & 0.722$\pm$0.001   &  -5.427$\pm$0.002  & 5.562$\pm$0.001   \\
$\rho_\star$ ($10^{-3}$)  & 2.64$\pm$0.01     & 2.36$\pm$0.01     & 2.37$\pm$0.01     &  2.38$\pm$0.01     & 2.36$\pm$0.01     \\
$\pi_{{\rm E},N}$         & -                 & -1.42$\pm$0.06    & 1.49$\pm$0.07     &  1.45$\pm$0.03     & -                 \\
$\pi_{{\rm E},E}$         & -                 & -0.34$\pm$0.04    & -0.19$\pm$0.06    &  -0.38$\pm$0.02    & -                 \\
$ds/dt$ yr$^{-1}$         & -                 &  -                &  -                &  0.05$\pm$0.01     & -                 \\
$d\alpha/dt$ yr$^{-1}$    & -                 &  -                &  -                &  -0.04$\pm$0.01    & -                 \\
$\xi_{{\rm E},N}$         & -                 &  -                &  -                &  -                 & -2.18$\pm$0.03    \\  
$\xi_{{\rm E},E}$         & -                 &  -                &  -                &  -                 & 0.19$\pm$0.11     \\
$\psi$ (deg)              & -                 &  -                &  -                &  -                 & 247.8$\pm$2.0     \\
$\ell$ (deg)              & -                 &  -                &  -                &  -                 & 10.3$\pm$1.9     
\enddata                                                                                     
\tablecomments{                                                            
HJD'=HJD-2450000. We note that the lensing parameters t0 and u0 are measured with respect to the center of the caustic located on the planet side.
}
\end{deluxetable*}

There exist several causes of long-term deviations in lensing light curves. The 
first is the change of the observer¡¯s position caused by the orbital motion of 
the Earth around the Sun \citep{gould92, alcock95}. This ``parallax effect'' 
causes the source trajectory to deviate from rectilinear, resulting in long-term 
deviations. The second is the positional change of the lens caused by the orbital 
motion of the binary lens \citep{dominik98, albrow00, bennett10, penny11, shin11, 
shin12, skowron11}. In addition to causing the source trajectory to deviate from 
rectilinear, the ``lens orbital effect'' causes further deviation in lensing light 
curves by deforming the caustic over the course of the event. The last cause of 
the deviation is the change of the source position caused by its orbital motion, 
if the source is a binary \citep{han97, dominik98}. Since this affects lensing 
light curves similarly to the parallax effect, it is often referred to as the 
``xallarap effect'', which is parallax spelled backward.

Considering that the parallax effect requires 2 parameters $\pi_{{\rm E},N}$ and 
$\pi_{{\rm E},E}$, that represent the two components of the lens parallax vector 
$\pivec_{\rm E}$ projected onto the sky along the north and east equatorial coordinates, 
respectively. The magnitude of the parallax vector corresponds to the relative 
lens-source parallax, $\pi_{\rm rel}= {\rm AU}(D_{\rm L}^{-1}-D_{\rm S}^{-1})$, 
scaled to the Einstein radius of the lens, i.e., $\pi_{\rm E}=\pi_{\rm rel}/\theta_{\rm E}$
\citep{gould04}. To first order approximation, the lens orbital motion is described 
by 2 parameters $ds/dt$ and $d\alpha/dt$ that represent the change rates of the 
normalized binary separation and the source trajectory angle, respectively 
\citep{albrow00}.  Modeling the xallarap effect requires 5 parameters: the 
components of the xallarap vector, $\xi_{{\rm E},N}$ and $\xi_{{\rm E},N}$, the 
orbital period $P$, inclination $i$ and the phase angle $\psi$ of the source orbital 
motion. The magnitude of the xallarap vector $\xivec_{\rm E}$ corresponds to the 
semi-major axis of the source's orbital motion with respect to the center of mass 
normalized by the projected Einstein radius onto the source plane \citep{dong09}.

\section{Result}

We test models considering the higher-order effects, and the results are summarized 
in Table~\ref{table:two}. From the comparison of the results, we find the following 
results. First, it is found that the parallax effect substantially improves the fit 
as shown by the residuals in Figure~\ref{fig:one}. We find that the improvement is 
$\Delta\chi^2\sim 752$ compared to the standard binary-lens model. Second, when we 
additionally consider the lens orbital effect, on the other hand, the improvement 
of the fit $\Delta\chi^2\sim 1$ is meager. Finally, we find that considering the 
xallarap effect yields solutions as good as the parallax solution for source orbital 
periods $P > 0.6$ yrs. This is expected because it is known that xallarap effects can 
mimic parallax effects \citep{smith03, dong09}. However, the xallarap solutions are 
excluded because they result in masses of the source companion bigger than 3 $M_\odot$ 
and this contradicts to the upper limit set by the observed blended light. Therefore, 
we conclude that the dominant effect for the long-term deviation is the parallax effect. 
Finally, since the source lies very near the ecliptic, it is subject to the 
``ecliptic degeneracy'', which has almost identical parameters except $(u_0, \alpha, 
\pi_{{\rm E},N}) \rightarrow -(u_0, \alpha, \pi_{{\rm E},N})$ \citep{skowron11}.

In Figure~\ref{fig:one}, we present the best-fit model (parallax model in 
Table~\ref{table:two}) curve that is overplotted on the observed light curve. 
In Figure~\ref{fig:two}, we also present the geometry of the lens system for the 
best-fit solution. It is found that the lens consists of binary components with a 
projected separation bigger than the Einstein radius corresponding to the total mass 
of the binary. For such a binary lens, there exist two sets of 4-cusp caustics, where 
one small set is located close to the heavier lens component (primary) and the other 
bigger set is located toward the lower-mass lens component (companion). The event 
was produced by the source trajectory passing the tips of the caustic located on 
the companion side. The strong peaks at HJD$\sim$2456537.5 and 2456542.5 were produced 
at the moments of the source crossings over the caustic tips, while the extended 
weak bump centered at HJD$\sim$2456065 was produced as the source passed through the 
magnification zone of the primary lens. Despite the relatively short time scale 
$t_{\rm E}\sim 26.5$ days of the event, clear detection of the parallax effect was 
possible due to combination of the large value of the lens parallax combined with 
the good coverage of the extended bump that continued for almost 2 months after 
the main peaks.

Detecting the parallax effect is important for the determinations of the physical 
lens parameters because the lens parallax $\pi_{\rm E}$ is related to the mass and the distance 
to the lens by $M_{\rm tot}=\theta_{\rm E}/(\kappa\pi_{\rm E})$ and $D_{\rm L} = 
{\rm AU}/(\pi_{\rm E}\theta_{\rm E}+\pi_{\rm S})$, respectively. Here $\kappa = 
4G/(c^2{\rm AU})$, $\pi_{\rm S} = {\rm AU}/D_{\rm S}$ is the parallax of the source 
star, and $D_{\rm S}$ is the distance to the source star. The source is in the Galactic 
bulge and thus its distance is known. Considering the mass distribution of the Galactic 
bulge and the projected source location, we estimate that $D_{\rm S} = 7.60$ kilo-parsecs, 
corresponding to $\pi_{\rm S}=0.132$ milli-arcseconds.

\begin{deluxetable}{lrr}
\tablecaption{Physical Parameters\label{table:three}}
\tablewidth{0pt}
\tablehead{
\multicolumn{1}{c}{parameter} &
\multicolumn{1}{c}{for $u_0>0$} &
\multicolumn{1}{c}{for $u_0<0$}
}
\startdata
total mass ($M_\odot$)                       &  0.024$\pm$0.002  &  0.024$\pm$0.002  \\
primary mass ($M_\odot$)                     &  0.023$\pm$0.002  &  0.022$\pm$0.002  \\
companion mass ($M_{\rm J}$)                 &  1.89$\pm$0.19    &  1.85$\pm$0.19    \\
projected separation (AU)                    &  0.89$\pm$0.03    &  0.87$\pm$0.03    \\
distance (kilo-parsec)                       &  1.79$\pm$0.12    &  1.76$\pm$0.13    \\ 
height above plane (parsec)                  &  106$\pm$7        &  106$\pm$7        \\
velocity, rotation direction (km s$^{-1}$)   &  -2$\pm$8         &  -59$\pm$8        \\
velocity, vertical direction (km s$^{-1}$)   &  17$\pm$6         &  -16$\pm$6        
\enddata  
\end{deluxetable}

\begin{figure}[ht]
\epsscale{1.2}
\plotone{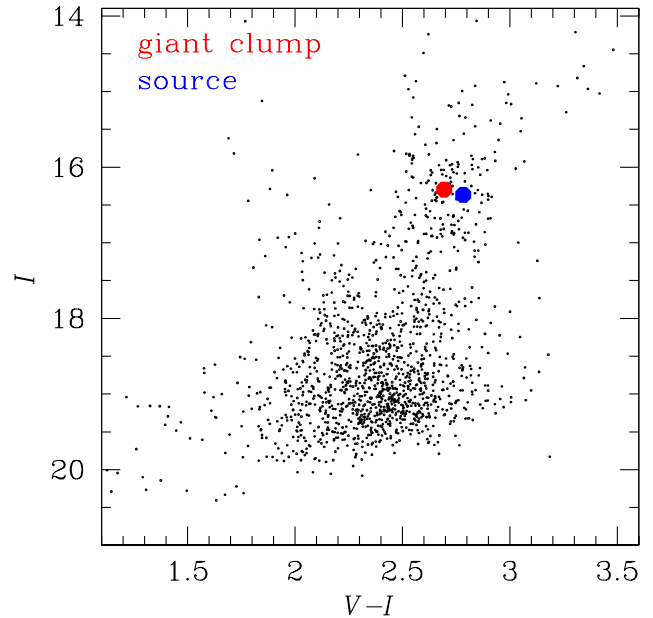}
\caption{\label{fig:three}
Location of the lensed source star in the color-magnitude diagram (marked in blue dot) 
relative to the centroid of the red clump (red dot). The magnitude and color are 
instrumental scale of the OGLE III data photometry.
}\end{figure}

For the full characterization of the physical parameters, it is needed to additionally 
determine the Einstein radius, which is given by $\theta_{\rm E}=\theta_*/\rho_*$. 
The normalized source radius $\rho_*$ is measured by analyzing the caustic-crossing 
parts of the light curve that are affected by finite-source effects. The angular source
radius $\theta_*$ is estimated from the source type that is determined based on its 
de-reddened color and brightness. For this, we first calibrate the color and brightness 
by using the centroid of bulge giant clump as a reference \citep{yoo04}, for which 
the de-reddened brightness $I_{0,{\rm c}}=14.45$ at the Galactocentric distance 
\citep{nataf13} and color $(V-I)_{0,{\rm c}}=1.06$ \citep{bensby11} are known. We 
then translate $V-I$ into $V-K$ color by using the color-color relations \citep{bessell88} 
and then find $\theta_*$ using the relation between the $V-K$ and the angular radius 
\citep{kervella04}. Figure~\ref{fig:three} shows the location of the source star in 
the color-magnitude diagram of stars in the same field obtained by the OGLE III experiment. 
It is found that the source is a K-type giant with an angular radius $\theta_* = 6.89 
\pm 0.60$ micro-arcseconds. The estimated Einstein radius is $\theta_{\rm E} = 0.29
\pm 0.03$ milli-arcseconds. Combined with the measured Einstein time scale $t_{\rm E}$, 
the relative lens-source proper motion is $\mu = \theta_{\rm E}/t_{\rm E} = 4.02 \pm 
0.37$ milli-arcseconds per year.

In Table~\ref{table:three}, we present the determined physical parameters of the lens. 
The mass of the companion is twice that of the Jupiter. The mass of the primary is 
$0.022 \pm 0.002\ M_\odot$. This is firmly below the hydrogen-burning limit of 
$0.08\ M_\odot$ and thus the primary is a BD. The lens is located at a distance 
$D_{\rm L}=1.76 \pm 0.13$ kilo-parsecs from the Earth toward the Galactic center. 
Then the projected separation between the lens components is $\ell_\perp = 
s D_{\rm L} \theta_{\rm E} = 0.87 \pm 0.03$ AU.

We also show the height above the Galactic plane $z$ and the tranverse velocity 
$(v_l, v_b)$ in the directions of Galactic rotation and Galactic north pole, 
respectively. To find the latter two, we measure the source proper motion 
$(\mu_N, \mu_E)_{\rm S} = (-0.20 \pm 0.65, 0.02 \pm 0.65)$ mas yr$^{-1}$ relative 
to the Galactic bar, and correct for the bar proper-motion gradient \citep{gould13}. These 
kinematic variables are the only ones that differ significantly between the two 
solutions resulting from the ecliptic degeneracy with $u_0 > 0$ and $u_0 < 0$. 
However, both sets of $(v_l, v_b)$ as well as $z$ are consistent with a lens age 
in the range 1 -- 10 Giga-years, i.e., much older than BDs of this mass found in 
imaging studies.

\section{Discussion}

The properties of the OGLE-2012-BLG-0358L system are relatively extreme compared 
to other binaries with BD hosts. In particular, the separation is a factor $\sim 15$ 
and $\sim 40$ times smaller than those of 2MASS 1207-3932 and 2MASS 0441-2301, 
respectively, and the mass ratio of is a factor $\sim 2$ and $\gtrsim 3$ times 
smaller than the mass ratios of these systems. Systems with such extreme properties 
may be difficult to form via conventional binary BD formation mechanisms (e.g, 
\citet{bate12}), suggesting an alternative scenario where the companion formed in 
the protoplanetary disk of the host BD. Surveys for disks around young BD have found 
some systems with inferred disk masses up to and even slightly exceeding $\sim M_{\rm J}$ 
\citep{harvey12}, although these are relatively rare and the inferred masses are 
subject to considerably uncertainty. Such massive disks are likely to be near the 
limit of stability (e.g., \citet{lodato05}), arguing for a gravitational instability 
formation scenario rather than core accretion. On the other hand, the relatively 
close separation may pose a challenge for gravitational instability. Clearly, 
additional theoretical work is needed to explore the viability of planet formation 
in BD protoplanetary disks, either by the gravitational instability or core accretion 
mechanism. For this, it is essential to find more binaries with BD hosts in wide 
ranges of mass ratios and separations.

Microlensing surveys for exoplanets are well-suited to detect planetary companions 
to very faint, low-mass stars and old BDs, systems which are difficult to discover 
via other methods. The last two decades have witnessed tremendous progress in 
microlensing experiments, which have enabled a nearly 10-fold increase in the 
observational cadence, resulting in an almost 100-fold increase in the event 
detection rate. With this observational progress, the number of BD events with 
precisely measured physical parameters is rapidly increasing \citep{shin13, choi13}. 
Furthermore, a new survey based on a network of multiple telescopes equipped with 
large format cameras is planned to achieve an even higher cadence of more than 100 
per day. Hence, starting from the system reported in this work, many additional 
BD hosts will be surveyed via microlensing. The discovery of additional, close 
separation, BD/planet systems with even more extreme mass ratio systems from these 
surveys will provide important empirical constraints on the ubiquity and mechanisms 
of planet formation around these hosts.

\acknowledgments 
Work by CH was supported by Creative Research Initiative Program (2009-0081561) 
of National Research Foundation of Korea. 
The OGLE project has received funding from the European Research Council under 
the European Community's Seventh Framework Programme (FP7/2007-2013) / ERC grant 
agreement no. 246678. 
The MOA experiment was supported by grants JSPS22403003 and JSPS23340064. TS 
acknowledges the support JSPS 24253004. TS is supported by the grant JSPS23340044. 
YM acknowledges support from JSPS grants JSPS23540339 and JSPS19340058.
AG and BSG acknowledge support from NSF AST-1103471. BSG, AG, and RWP acknowledge 
support from NASA grant NNX12AB99G. SDong¡¯s research was performed under contract 
with the California Institute of Technology funded by NASA through the Sagan Fellowship 
Program. 
KA, DB, MD, KH, MH, SI, CL, RS, YT are supported by NPRP grant NPRP-09-476-1-78 from 
the Qatar National Research Fund (a member of Qatar Foundation). 
MD is a Royal Society University Research Fellow. KH is a Royal Society Leverhulme 
Trust Senior Research Fellow. CS received funding from the European Union Seventh 
Framework Programme (FP7/2007-2013) under grant agreement no. 268421. The research 
leading to these results has received funding from the European Community's Seventh 
Framework Programme (/FP7/2007-2013/) under grant agreement no. 229517.

\end{document}